\documentclass[12pt]{iopart}
\usepackage{graphicx}
\usepackage{iopams} 

\begin{document}

\title[Fluctuations and entanglement spectrum in quantum Hall states]{Fluctuations and entanglement spectrum in quantum Hall states \footnote{Corresponding author karyn.lehur@cpht.polytechnique.fr. This paper is dedicated to the JSTAT special issue on Quantum Entanglement in Condensed Matter Physics.}}

\author{Alexandru Petrescu$^{1,2}$, H. Francis Song$^{3}$, Stephan Rachel$^4$, Zoran Ristivojevic$^{2}$, Christian Flindt$^{5}$, Nicolas Laflorencie$^{6}$, Israel Klich$^{7}$, Nicolas Regnault$^{8,9}$, Karyn Le Hur$^{2}$}

\date{\today}

\address{$^1$ Department of Physics, Yale University, New Haven, CT 06520}
\address{$^2$ Centre de Physique Th\'eorique, Ecole Polytechnique, CNRS, 91128 Palaiseau Cedex France} 
\address{$^3$ Center for Neural Science, New York University, New York, NY 10003, USA}
\address{$^4$ Institute for Theoretical Physics, TU Dresden, 01062 Dresden, Germany} 
\address{$^5$ D\'epartement de Physique Th\'eorique, Universit\'e de Gen\`eve, CH-1211 Gen\`eve, Switzerland} 
\address{$^6$ Laboratoire de Physique Th\' eorique, Universit\' e de Toulouse, UPS, (IRSAMC), Toulouse, France} 
\address{$^7$ Department of Physics, University of Virginia, Charlottesville, VA 22904} 
\address{$^8$ Department of Physics, Princeton University, Princeton, NJ 08544}
\address{$^9$ Laboratoire Pierre Aigrain, ENS-CNRS UMR 8551, Universit\'es P. et M. Curie and Paris-Diderot, 24, rue Lhomond, 75231 Paris Cedex 05, France}

\begin{abstract}
The measurement of quantum entanglement in many-body systems remains challenging. One experimentally relevant fact about quantum entanglement is that in systems whose degrees of freedom map to free fermions with conserved total particle number, exact relations hold relating the Full Counting Statistics associated with the bipartite charge fluctuations and the sequence of R\' enyi entropies. We draw a correspondence between the bipartite charge fluctuations and the entanglement spectrum, mediated by the R\' enyi entropies. In the case of the integer quantum Hall effect, we show that it is possible to reproduce the generic features of the entanglement spectrum from a measurement of the second charge cumulant only. Additionally, asking whether it is possible to extend the free fermion result to the $\nu=1/3$ fractional quantum Hall case, we provide numerical evidence that the answer is negative in general. We further address the problem of quantum Hall edge states described by a Luttinger liquid, and derive expressions for the spectral functions of the real space entanglement spectrum at a quantum point contact realized in a quantum Hall sample. 
\end{abstract}


\section{Introduction}
Quantum information concepts have become essential in the study of condensed-matter systems and quantum phase transitions. 
Traditionally quantum order has been defined through the study of correlation functions. The latter distinguish different phases around a critical point. An alternative quantity describing correlations is the \textit{entanglement entropy} $S$ \cite{vN1955}. It follows from the reduced density matrix corresponding to the partition of a system into two complementary subsystems. Criticality and the degree of quantum entanglement of a pure quantum state are indicated by the scaling of $S$ with the linear dimension of the subsystem $\ell$. That is, $S$ can scale with the area of the subsystem boundary $S \sim \ell^{ d - 1 }$ \cite{KP2006, LW2006, ECP2010} for systems in a $d-$dimensional volume. This is true in gapped systems \cite{WEA2008}, where intuitively due to the finite correlation length only modes close to the subsystem boundary contribute to entropy. There are however critical models which exhibit area laws \cite{CEA2006,VEA2006,MFS2009}. Multiplicative logarithmic corrections can occur, for example for gapless lattice fermions \cite{AreaLaws} and in Heisenberg antiferromagnets \cite{ln}. In critical models which can be mapped to a conformal field theory, the entanglement entropy scales logarithmically with subsystem size $S \sim \log \ell$ \cite{HLW1994,CC2009}. Topological order \cite{WenNiu1990} in the gapped ground state of some two-dimensional systems can be described by a universal subleading term in the entanglement entropy \cite{KP2006, LW2006}, known as topological entanglement entropy. The entanglement entropy is a single number characterizing the full set of eigenvalues of the reduced density matrix, the \textit{entanglement spectrum} \cite{LH,thomale,papic,haque}. Recently, it has been suggested by Li and Haldane \cite{LH} that the entanglement spectrum yields a more complete description of topological order. They have showed that the low-lying entanglement spectrum resembles the edge excitation spectrum and thus can be used to distinguish topological orders.

The entanglement entropy can arise from an effective thermodynamic description of the quantum ground state.  Li and Haldane recast the reduced density matrix as $\exp(-H_E)$ so that the entanglement entropy becomes equivalent to the canonical ensemble entropy of a system described by a Hermitian \textit{entanglement Hamiltonian} $H_E$ at an effective entanglement temperature $T_E=1$. Alternative definitions of such effective thermodynamics at zero temperature might be possible based on a temperature that is non-universal and essentially depends on the coupling between the two subsystems of the many-body quantum system \cite{ELR2006, Wetal2011}. 

For all its deep theoretical implications, the measurement of quantum entanglement remains challenging beyond the case of a few entangled objects \cite{Aspect,Hagley}. Some efforts have been done to relate thermodynamical quantities and entanglement properties in real materials \cite{Gegenwart,Zeilinger}.The entropy can be measured in the case of relatively simple systems where one subsystem can be identified to a spin-1/2 particle \cite{KoppLeHur,Karyn,Affleck}. More generally, the entropy can be obtained from the population of low lying energy levels after a local quantum quench \cite{C2011,StephanDubail}, or the occupation of the states of the quantum switch operating the quench \cite{AD2012}. Ultracold atom measurement protocols based on similar concepts have been proposed in Refs. \cite{Detal2012, Petal2013}. The flow of R\'enyi entropies in quantum transport has been discussed \cite{N2011}. Moreover, the entanglement entropy could be obtained from the Shannon entropy of the probability distribution of certain symmetry observables \cite{KRS2006}. One experimentally relevant manifestation of quantum entanglement is that in systems with a conserved $U(1)$ current charge within the subsystem can fluctuate. Recently, Klich and Levitov \cite{KL2009} and some of us \cite{Setal, SetalR} have suggested to focus on the concept of {\it bipartite fluctuations} and more generally on the {\it full counting statistics} (FCS) emerging in many-body quantum systems when tracing out one of the two blocks of a spatial bipartition. Bipartite fluctuations can also be used to detect quantum phase transitions \cite{quantumfluc}. Whenever the system can be mapped to free fermions with conserved total particle number $N$, exact relations hold between FCS and the sequence of R\'enyi entropies \cite{R1961}. Mediated by the R\'enyi entropies, a correspondence between the entanglement spectrum and FCS follows. In the general case of interacting fermions, similar expressions are not expected to hold \cite{HGF2009, FK2013}. For free bosons, similar relations can also be applied \cite{Nataf}.

We show in this work that under common experimental conditions it is possible to obtain good approximations to the sequence of R\'enyi entropies from only the second charge cumulant (charge noise).  R\'enyi entropies themselves can be used to extract the density of levels of the low lying entanglement spectrum. Since bipartite fluctuations and higher order charge cumulants have been measured in mesoscopic condensed-matter systems \cite{Glattli,Picciotto,meso1,meso2,meso3} and in cold atomic gases \cite{atoms1,atoms2}, this is a feasible route to the measurement of quantum entanglement.

The remainder of this paper is organized as follows. In Sec.~\ref{Sec:MainEquations}, we begin with a general discussion of the relation between FCS and the R\'enyi entropies in free fermion systems with conserved charge, drawing on previous work \cite{KL2009, Setal}. We afterwards prove in Sec.~\ref{Sec:bulkIQHE} that for bulk states in an integer quantum Hall sample \cite{Pepper} it is possible to reproduce the generic features of the entanglement spectrum from a measurement of the second charge cumulant only.  We next turn to the fractional quantum Hall case in Sec.~\ref{Sec:FQHE}. This strongly correlated state falls beyond the scope of the exact relations used so far. We exemplify this on the Laughlin state \cite{Laughlin}. In Sec.~\ref{Sec:1DModels} we address the problem of quantum Hall edge states described by Luttinger theory \cite{HaldaneL,G2003}, by focusing on geometries with a quantum point contact that naturally separates the system into two blocks.

\section{Real space bipartite fluctuations and the entanglement spectrum} 
\label{Sec:MainEquations}
We consider a physical system described by the zero temperature pure ground state density matrix $\rho = |\psi\rangle \langle \psi |$. We assume a bipartition of the physical degrees of freedom into complementary subsets $A$ and $B$. The reduced density matrix of subsystem $A$ is obtained by tracing the degrees of freedom belonging to $B$,
\begin{equation}\label{eq:rA}
\rho_A = \mathrm{Tr}_B \rho.
\end{equation}
The reduced density matrix is normalized $\mathrm{Tr}_A \left( \mathrm{Tr}_B \rho \right) = 1$. It no longer represents a pure state, owing to the entanglement between the subsystems. We will mostly consider situations where $A$ and $B$ form a real space partition (RSP).  One measure for the entanglement between part $A$ and part $B$ is the von Neumann entanglement entropy \cite{vN1955} defined as
\begin{equation}
\label{Eq:S}
S = - \mathrm{Tr}_A \left(\rho_A \log \rho_A\right).
\end{equation}
While the von Neumann entanglement entropy $S$ is a number characterizing the set of eigenvalues of $\rho_A$, their spectral function can be deduced from the sequence of R\'enyi entropies $\{ S_n \;| \; n \geq 1 \}$ \cite{R1961}. The von Neumann entropy of Eq.~(\ref{Eq:S}) is recovered in the limit $n\to 1$ from the R\'enyi entropy $S_n$, where
\begin{equation}
\label{Eq:Sn}
S_n= \frac{1}{1-n} \log \mathrm{Tr}_A\,(\rho_A)^n.
\end{equation}

We define the \textit{entanglement spectrum} (ES) as the set $\{\lambda_i\}$ of eigenvalues of $\rho_A$. In the rest of the paper, we characterize the ES through the spectral function $P(\lambda) = \sum_i \delta( \lambda - \lambda_i )$. We can relate $P(\lambda)$ to the R\'enyi entropies. For this we introduce the sequence of traces of powers of $\rho_A$
\begin{equation}
\label{Eq:Rn}
R_n =\sum_i \lambda_i^n = \exp\left[( 1 - n ) S_n \right],\; n \geq 1.
\end{equation}
The spectral function is then determined uniquely \cite{CL2008} by $P(\lambda) = \frac{1}{\lambda} \lim_{\epsilon \rightarrow 0} \mathrm{Im}~ F(\lambda - i \epsilon)$, where $F( z )$ is the function of one complex variable
\begin{equation}
\label{Eq:fz}
F(z) = \frac{1}{\pi}  \sum_{n=1}^\infty \frac{R_n}{z^n} = \frac{1}{\pi} \int_0^1 d\lambda \frac{\lambda P(\lambda)}{z-\lambda}.
\end{equation}
In the following, we will employ the counting function
\begin{equation}
\label{Eq:nlambda}
n( \lambda ) = \int_{\lambda}^{\lambda_\mathrm{max}} d \tilde{ \lambda }~P( \tilde{ \lambda } ).
\end{equation}
If $\lambda_\mathrm{max}$ is the maximum eigenvalue of the ES, then $n( \lambda )$ is the number of levels in the ES enclosed between $\lambda$ and $\lambda_\mathrm{max}$.

The ES spectral function can be obtained from the formal equality of Eq.~(\ref{Eq:fz}) whenever the traces $R_n$ are known to good approximation. As we will exemplify, one effective approximation can be devised in noninteracting fermionic systems, where exact relations exist between the sequence of R\'enyi entropies $S_n$ and that of bipartite charge cumulants $C_m$ \cite{Setal}. The latter arise by differentiation of the cumulant generating function
\begin{equation}\label{Eq:Cn}
C_m=(-i\partial_\lambda)^m\ln\langle \exp(i\lambda N_A)\rangle|_{\lambda=0},
\end{equation}
where $N_A$ is the number of particles in the subsystem $A$, and the expectation value is taken in the many body ground state $|\psi\rangle$. The entanglement entropy and the R\'enyi entropies in free particle systems are related to charge fluctuations in the subsystem $A$. The former can be expressed as
\begin{equation}
\label{Eq:Sse}
S = \lim_{M\rightarrow \infty} \sum_{m=1}^{M+1}\alpha_{m}(M) C_{m},
\end{equation}
where
\begin{equation}
\label{Eq:a}
\alpha_{m}(M) = \left\{
\begin{array}{ll}
  2\sum_{r=m-1}^M \frac{S_1(r,m-1)}{r! r}, &  m\; \mathrm{ even,} \\
  0, &  m\; \mathrm{ odd.}
  \end{array}
\right.
\end{equation} 
$S_1(k,n)$ denotes the unsigned Stirling number of the first kind \cite{GR}.  Similarly, the set of R\'enyi entropies can be represented as \cite{Setal}
\begin{equation}
\label{Eq:Snse}
S_n = \lim_{M\rightarrow \infty} \sum_{k=1}^{n M} \beta_{k}(n,M) C_{k},
\end{equation}
where $\beta_{k}(n, M )$ are numerical coefficients independent of the Hamiltonian \cite{Setal}. Importantly, the convergence properties of Eqs.~(\ref{Eq:Sse}) and (\ref{Eq:Snse}) can be exploited to use truncated series of cumulants as approximations to exact entropies.

In practice, the series of Eqs.~(\ref{Eq:Sse}) and ~(\ref{Eq:Snse}) can be further simplified in cases where only the second cumulant is nonvanishing \cite{Cumulants}, or if the higher order cumulants are suppressed, which is the case at high particle density \cite{CMV2012}. In Eq.~(\ref{Eq:Sse}) the limit $M\rightarrow \infty$ can be commuted with the summation of the series, leading to the simpler forms $\lim_{ M \rightarrow \infty} \alpha_{2m} ( M ) = s^{(1)}_{2m}$. For the coefficients of Eq.~(\ref{Eq:Snse}), $\lim_{M \rightarrow \infty} \beta_{2m}( n, M ) = s^{(n)}_{2m}$, the latter being the coefficient of the $2m$-th cumulant in the series for $S_n$ \cite{KL2009,CMV2012}
\begin{equation}
\label{Eq:sal}
s^{(n)}_{2m} = \frac{(-1)^n (2\pi)^{2m} 2 \zeta\left[ -2m, \left(1+n\right)/2 \right] }{(n - 1) n^{2m} (2m)! }.
\end{equation}
Here $\zeta$ is the generalized Zeta function \cite{GR}.

In this work we study two cases where essentially only the second cumulant, $C_2$, is needed to accurately truncate the series of Eqs.~(\ref{Eq:Sse}) and (\ref{Eq:Snse}). Firstly, for a non-interacting Fermi gas of $N$ particles in a $d$-dimensional volume, in the limit $N\to\infty$ at fixed volume, the asymptotic behavior of the R\'enyi entropy $S_n$ is given by the second cumulant, i.e. 
\begin{equation}
\frac{S_n}{C_2} \to s^{(n)}_2\;\mathrm{as}\; \frac{N}{L^d} \to \infty   
\end{equation}
with $s^{(n)}_2$ defined in Eq.~(\ref{Eq:sal}). This result was proved in Ref.~\cite{CMV2012}, which uses the fact that, for free fermions, both $S_n$ and $C_2$ have leading order terms of the form $N^{(d-1)/d} \log N$ as $N \to \infty$ at constant volume $L^d$, whereas this same leading order contribution vanishes in higher order cumulants $C_n$ for $n \geq 3$.  Since entropies are accessible from $C_2$ under the feasible assumption of high density, we can determine the counting $n(\lambda)$ of the entanglement spectrum in the integer quantum Hall effect in Sec.~\ref{Sec:bulkIQHE}. Secondly, we address the case of a gaussian theory, where by definition $C_2$ is the only nonvanishing cumulant. Mapping the edge subsystem of an integer quantum Hall bar to a Luttinger liquid, we revisit the following physical situation: the quantum Hall sample is pinched off with a strong gating voltage such as to create a ``weak link'' at which edge electrons can be transfered via a weak tunneling term. In Sec.~\ref{Sec:WLt}, we perturbatively calculate the entanglement entropy build-up in one half of the sample as a function of observation time, and obtain the corresponding time dependent counting function $n( \lambda; t )$.

\section{Entanglement measures from charge fluctuations in the quantum Hall effect}
We begin in Subsec.~\ref{Sec:bulkIQHE} with the example of the bulk integer quantum Hall effect at $\nu = 1$. In a certain limit accessible to experiment, the R\'enyi entropies are determined by $C_2$. From this result, we are able to reconstruct the counting function $n(\lambda)$ [Eq.~(\ref{Eq:nlambda})] in good agreement with exact numerical calculation. We secondly show in Subsec.~\ref{Sec:FQHE} that the entanglement entropy of the $\nu = 1/3$ Laughlin state is not well approximated by cumulant series Eq.~(\ref{Eq:Sse}).

\subsection{Entanglement spectrum in the integer quantum Hall effect at $\nu = 1$}
\label{Sec:bulkIQHE}
We consider a disk of total radius $R$, and take the magnetic length in SI units to be $l = \sqrt{ \hbar c / ( e B ) }$. In the symmetric gauge, the lowest Landau level (LLL) eigenfunctions for angular momentum $m \geq 0$ are given by
\begin{equation}
\label{Eq:phim}
\phi_m(z) = \frac{(z/l)^m e^{-\frac{1}{4}|z/l|^2}}{ l \sqrt{2\pi 2^m m!}}.
\end{equation}
We denote by $z = x + i y$ the complex coordinate in the plane. The normalized probability distribution $|\phi_m(z)|^2$  is concentrated near a circle of radius $r=\sqrt{2m}\; l$. Since the system has radius $R$, finitely many orbitals are supported. The angular momentum of an LLL particle can take one of $m_\mathrm{max}+1$ values in the set of orbitals $\mathcal{O} = \{0,...,m_\mathrm{max}\}$, where $m_\mathrm{max} = \frac{R^2}{2l^2}$. Therefore the LLL wavefunction is
\begin{equation}
\label{Eq:LLL}
| \Psi_\mathrm{LLL} \rangle = \prod_{m \in \mathcal{O}} c_m^\dagger | 0 \rangle,
\end{equation}
where $| 0 \rangle$ is the state with no fermions. 

We obtain the real space partition entanglement spectrum (RSP ES) for an annular subsystem $A$ of inner radius $r_1 \geq 0$ and outer radius $r_2 > r_1$. The reduced density matrix is related to the Green's function matrix $G_{ij} = \langle \Psi_\mathrm{LLL} |  c_i^\dagger  c_j | \Psi_\mathrm{LLL} \rangle$ \cite{P2003,RS2009}. Since the cut is invariant to azimuthal rotation, the restriction of $G$ to $i, j \in A$ is block diagonal with respect to angular momentum. The eigenvalue corresponding to some orbital $m \in \mathcal{O}$ is
\begin{equation}
\label{Eq:zetam}
\zeta_m = \int_A dz |\phi_m(z)|^2 = \frac{1}{m!}\Gamma\left(m+1, \frac{r_1^2}{2l^2},\frac{r_2^2}{2l^2}\right).
\end{equation}
The associated eigenvector is $\phi^*_m(z)$. In Eq.~(\ref{Eq:zetam}), $\Gamma$ is the incomplete Gamma function \cite{GR}. We note that $\zeta_m = 1$ if the orbital $\phi_m$ is localized well inside $A$, and 0 if it resides well outside $A$. In terms of the $\zeta_m$, the reduced density matrix $\rho_A = \mathrm{Tr}_B \rho$ is a $2^{m_\mathrm{max}+1} \times 2^{m_\mathrm{max}+1}$ dimensional matrix
\begin{figure}[t!]
\includegraphics[width=\linewidth]{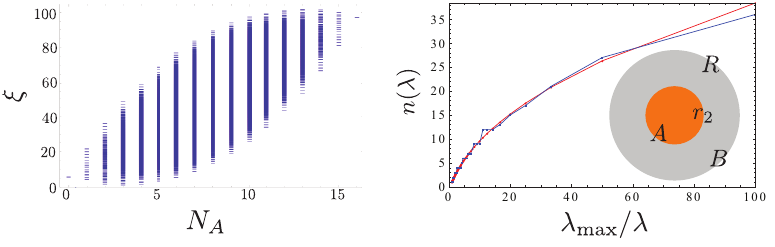}
\caption{\label{fig_rspes_iqh} RSP ES for $N = 16$ fermions in the lowest Landau level in a disk geometry: Subsystem $A$ is a disk of radius $r_2 = \sqrt{6}~l$, the entire system is a disk of radius $R=\sqrt{30}~l$ (see the inset of the right panel). \textbf{Left:}  Entanglement spectrum as pseudoenergies $\{ \xi_i \}$ versus total particle number $N_A$ in $A$. \textbf{Right:} For the spectrum in the left panel, approximate counting function $n(\lambda)$ obtained from $C_2$ (red) and computed $n(\lambda)$ (blue) of the RSP ES Eq.~(\ref{Eq:xiRSP})~[Note that $\lambda_\mathrm{max} = \exp( - \xi_\mathrm{min})$].}
\end{figure}

\begin{equation}
\label{Eq:rhoAzetam}
\rho_A = \bigotimes_{m \in \mathcal{O}} \mathrm{diag}(\zeta_m, 1 - \zeta_m).
\end{equation}
Then the von Neumann and R\'enyi entropies are
\begin{equation}
\label{Eq:SC}
S = \sum_{m \in \mathcal{O}} H_2 ( \zeta_m ), 
\;\; S_n = \sum_{m \in \mathcal{O}} \frac{1}{1-n} \log \left[ \zeta_m^n + (1-\zeta_m)^n \right],
\end{equation}
with $n \geq 2$. We have introduced the binary entropy function $H_2( x ) = -x \log x - ( 1 - x ) \log( 1 - x )$. Only orbitals close to the entanglement boundary $m \approx r_1^2 /( 2 l^2 )$ or $m \approx r_2^2 / ( 2 l^2 )$ participate in the expressions of the entanglement entropies, hence these are expected to scale with the perimeter of the boundary.

The RSP ES can be deduced directly from Eq.~(\ref{Eq:rhoAzetam}), which is a relation between $G_{ij}$ and $\rho_A$ \cite{P2003}. We take a slightly different route to the ES. By using Eq.~(\ref{Eq:SC}) together with Eq.~(\ref{Eq:fz}), we recover the ES from $S_n$, with $n \geq 1$. Noting that $R_n = 1$ for $n=1$ and $R_n = \prod_{m \in \mathcal{O}} \left[ \zeta_m^n + (1-\zeta_m)^n \right]$ for all $n > 1$, we find 
\begin{equation}
F( z ) = \frac{1}{\pi} \sum_{\{n_m\}} \frac{\lambda_{\{n_m\}}}{z-\lambda_{\{n_m\}}},
\end{equation}
where $\{n_m\}$ denote all sets of $m_\mathrm{max} + 1$ occupation numbers of the orbitals in the set $\mathcal{O}$. Each occupation number $n_m$ can be either 0 or 1. There are in total $2^{m_\mathrm{max}+1}$ sets of occupation numbers $\{ n_m \}$. To each set of occupation numbers $\{ n_m \}$ one associates a level in the entanglement spectrum 
\begin{equation}
\lambda_{\{n_m\}} \equiv \prod_{m \in \mathcal{O}} \zeta_m^{n_m} ( 1 - \zeta_m )^{ 1 - n_m }.
\end{equation} 
We define $\lambda_{\mathrm{max}}$ as the maximum of this set, which can be degenerate. The RSP ES can be recast in terms of the eigenvalues of the ``entanglement Hamiltonian'' $H_E$ discussed in the introduction
\begin{equation}
\label{Eq:xiRSP}
\xi_{\{n_m\}} = \xi + \sum_{ m \in \mathcal{O} } n_m \epsilon_m.
\end{equation}
We have introduced $\xi = - \sum_{ m \in \mathcal{O} } \log ( 1 - \zeta_m )$ and $\epsilon_m = \log( 1 - \zeta_m ) - \log( \zeta_m )$. In the left panel of Fig. \ref{fig_rspes_iqh}, the RSP ES in the form of Eq.~(\ref{Eq:xiRSP}) is plotted as a function of the total occupancy $N_A = \sum_{m = 0}^{m_\mathrm{max}} n_m$. Note that $\xi_\mathrm{min} = - \log \lambda_\mathrm{max}$.

We remark that Eq.~(\ref{Eq:xiRSP}) can be obtained by explicit Schmidt decomposition \cite{DRR2012}. Let $\langle z | c^\dagger_{Am} | 0 \rangle$ be a normalized wavefunction obtained from the restriction of $\phi_m(z)$ to $A$ (and similarly for $B$). Then the decomposition $c^\dagger_m = \zeta_m^{1/2} c_{Am}^\dagger + ( 1 - \zeta_m )^{1/2} c_{ Bm }^\dagger$ for each $m \in \mathcal{O}$ leads to the Schmidt decomposition of the many body state
\begin{equation}
\label{Eq:LLLd}
| \Psi_\mathrm{LLL} \rangle = \sum_{\{ n_m \}} \exp( - \xi_{\{ n_m \}} / 2 ) | \psi_A^{\{ n_m \}} \rangle \otimes |\psi_B^{\{n_m\}} \rangle.
\end{equation}
Note that the wavefunction $|\psi_A^{\{n_m\}}\rangle = \prod_{m = 0}^{m_\mathrm{max}} (c_{Am}^\dagger)^{n_m} | 0 \rangle$ is orthogonal by construction to $|\psi_B^{\{n_m\}}\rangle = \prod_{m = 0}^{m_\mathrm{max}} (c_{Bm}^\dagger)^{1 - n_m} | 0 \rangle$.

\begin{figure}[t!]
\includegraphics[width=\linewidth]{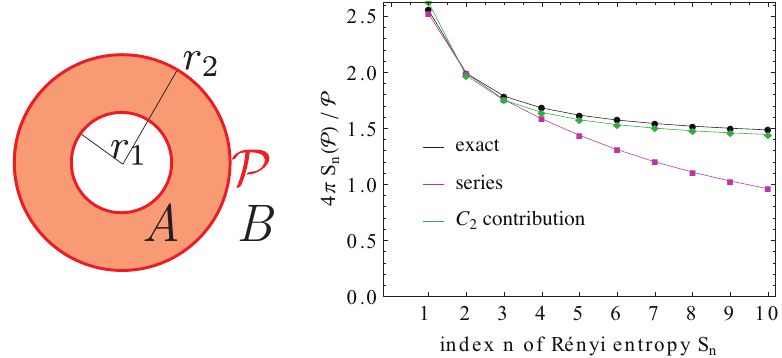}
\caption{\label{fig_rspee_iqh}
Entanglement entropies for an annular subsystem in the high density limit. \textbf{Left:} The annulus $A$, enclosed between $r_1$ and $r_2 = r_1 + 10~l$ in an infinite disk, supports a few hundred orbitals when $r_1 \sim 30~l$. \textbf{Right:} For $r_1 \gtrsim 20~l$, entropies obey a boundary law $S_n (\mathcal{P}) \propto \mathcal{P}$, where $\mathcal{P} = 2\pi (r_1 + r_2)$. We plot $4 \pi S_n(\mathcal{P}) / \mathcal{P}$ obtained in three ways: \textit{black circles} show the exact value obtained from the correlation matrix, \textit{magenta squares} show the approximation of $S_n$ via series [$M = 32$ for Eq.~(\ref{Eq:Sse}) and $ M = 5 $ for Eq.~(\ref{Eq:Snse})], \textit{dark green rhombi} show the approximation $S_n \approx s^{(n)}_2 C_2$.
}
\end{figure}

We next employ Eqs.~(\ref{Eq:Sse}) and (\ref{Eq:Snse}) to approximate entanglement entropies and thereby $P(\lambda)$. The cumulant generating function $\log \chi(\lambda)$ can be related to $G_{ij}$ with $i,j \in A$ via
\begin{equation}
\log \chi(\lambda) = \log \det \left[ 1 + ( e^{i \lambda} - 1 ) G \right].
\end{equation}
The first $10$ R\'enyi entropies, $S_1 = S$, $S_2,..., S_{10}$ are plotted on the right panel of Fig.~\ref{fig_rspee_iqh}, along with partial sums of cumulant series. $S_n$ for $n \geq 1$ and $C_{n}$ for $n \geq 1$ depend linearly on the entanglement perimeter. This is a manifestation of the ``boundary law'' in gapped fermion systems. In such cases, $S \propto \ell^{d-1}$ where the subsystem of volume $\ell^d$ is embedded in a $d-$dimensional system. 

Two remarks follow from the numerical study of the relation between entanglement entropies and cumulants in an annular subsystem of a disk (Fig.~\ref{fig_rspee_iqh}): Firstly, the number of terms required for the convergence of the series increases with the index $n$ of $S_n$. Secondly, only the first term $s^{(n)}_2 C_2$ approximates $S_n$ within a relative error of at most $3\%$ for the chosen annular geometry, which supports on the order of a hundred orbitals. This occurs since $C_2$ provides a leading order contribution in $N$ to the entropies as the number of particles is increased at finite volume $N / L^d \to \infty$ \cite{CMV2012}. In IQHE experiments this condition is achieved by increasing the $B-$field. At filling $\nu = 1$, the typical electron number is $N \sim 10^{11} B[T] \times R^2[cm^2]$. (Alternatively, increasing the fermion number at fixed $B-$field occupies higher Landau levels).

We can now approximate $n(\lambda)$ from $C_2$. We test the validity of the approximation by considering $N = 16$ fermions in the LLL. Even for such a comparably small particle number, $S_n$ are well approximated by the $C_2$ contribution. Approximate $R_n$ are
\begin{equation}
\label{Eq:RnApprox}
R_n \approx \exp \left[\frac{\pi^2}{6} \left(\frac{1}{n} - n\right) C_2 \right].
\end{equation}    
Then Eq.~(\ref{Eq:fz}) gives $n(\lambda)$, which is in good agreement with the actual counting function computed from the RSP ES of Eq.~(\ref{Eq:xiRSP}) (see Fig.~\ref{fig_rspes_iqh}).  In \ref{Ap:esiqhe} we provide criteria for the validity of the approximation in Eq.~(\ref{Eq:RnApprox}). The surface areas of subsystems $A$ and $B$ have to be approximately equal, and large compared to $2 \pi l^2$. The latter is intuitively the surface area on which most of the weight of an LLL orbital $\phi_m(z)$, for some orbital quantum number $m$, is enclosed. This is consistent with the requirement that many particles be accommodated in the finite area of the system \cite{CMV2012}.

We note that $C_2$ of a two-dimensional electron gas can be derived from measurements of the compressibility in the sub-region A, $\partial \langle N_A \rangle / \partial \mu_A$. The local compressibility in quantum Hall samples can be accessed, for example, with subsurface charge accumulation imaging \cite{T1998,GEA2001}. In a similar manner, in a spin system such as a two-dimensional antiferromagnet \cite{ln}, the spin susceptibility $\chi = \partial \langle S^z_A \rangle/ \partial h_A |_{h_A \to 0}$ yields the fluctuations in the total magnetization of $A$, $\langle S^z_A \rangle$. Magnetic flux is restricted to subsystem $A$ by screening the field $h_B$ over subsystem $B$ with superconducting Meissner screens \cite{Setal,ln}.

\subsection{Entanglement entropy in the fractional quantum Hall effect at $\nu=1/3$}
\label{Sec:FQHE}
While Eq.~(\ref{Eq:Sse}) and (\ref{Eq:Snse}) are true for free fermion systems, it is compelling to test them in a strongly correlated state. In this section, we compare the series result for the entanglement entropy with the exact result in the $\nu = 1/3$ Laughlin state \cite{Laughlin}. We find that the free fermion result is inapplicable in general. We study few fermion Laughlin wavefunctions in \ref{Ap:FQHE}. 

The computation of the ES for Laughlin states was performed in Ref.~\cite{S2012}. Here we compute the entanglement entropy for the representation of Laughlin states on the sphere \cite{H1983}. We consider a magnetic monopole at the center of the sphere such that the flux through the sphere is $N_\Phi$ flux quanta. The single particle orbitals are 
\begin{equation}
\label{Eq:phims}
\phi_m ( \mathbf{r} ) = \sqrt{\frac{ ( N_\Phi + 1 )! }{ 4 \pi m ! ( N_\Phi - m )!  }} u^m v^{ N_\Phi - m }. 
\end{equation}
Note that the particle coordinate $\mathbf{r}$ is parametrized by spinor coordinates $u=\cos(\theta/2)e^{i \varphi/2}$ and $v=\sin(\theta/2) e^{-i\varphi/2}$, corresponding to spherical polar angles $(\theta,\phi)$. The angular momentum along $z$ corresponding to Eq.~(\ref{Eq:phims}) is $L_z = N_\Phi / 2 - m$, where $m = 0, 1, ..., N_\Phi$.  Orbitals on the sphere in Eq.~(\ref{Eq:phims}) can be related to orbitals in the plane in Eq.~(\ref{Eq:phim}) via a stereographic projection $z \simeq u / v$, where the equality is up to a scaling factor. We therefore use complex $z$ to specify particle coordinates. We take the magnetic length $l$ to be unity.

The Laughlin state of $N$ particles at filling factor $\nu = 1 / m$ can be expressed in first quantized form as
\begin{equation}
\label{Eq:Laughlin}
\Psi_\mathrm{Laughlin}^{1/m} (z_1, ..., z_N ) \propto \prod_{ 1 \leq i < j \leq N } ( z_i - z_j )^m .
\end{equation}
For fermions, antisymmetry with respect to the interchange of any two coordinates $z_i$, $z_j$ requires that $m$ be an odd integer. We have suppressed from Eq.~(\ref{Eq:Laughlin}) the factor $e^{- \frac{1}{4 } \sum_{i=1}^{N} |z_i|^2}$ and the normalization factor. The case of $m=1$ corresponds to the Slater determinant of Eq.~(\ref{Eq:LLL}). Moreover, Eq.~(\ref{Eq:Laughlin}) constrains $N_\Phi = m(N-1)$, such that $N/N_\Phi = \nu$ in the thermodynamic limit. 

\begin{figure}[t!]
\includegraphics[width=\linewidth]{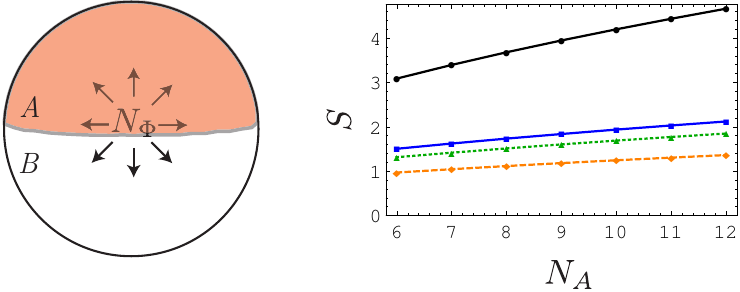}
\caption{\label{fig_rspee_fqh}
\textbf{Left:} A magnetic monopole threads $N_\Phi$ flux quanta through the surface of the sphere. The real space equatorial cut yields complementary hemispheres $A$ and $B$. \textbf{Right:} The entanglement entropy $S$ for the equatorial cut is represented by \textit{black circles}, versus the partial sum of the cumulant series of Eq.~(\ref{Eq:Sse}) with terms up to $C_{20}$, where the sum has converged (\textit{blue squares}), terms up to $C_4$ (\textit{green triangles, dashed line}), and only the contribution from $C_2$ (\textit{orange rhombi, long-dashed line}), in terms of the number of fermions in subsystem $A$.}
\end{figure}

To compute the RSP entanglement entropy we perform a real space equatorial cut \cite{S2012,DRR2012,RSS2012}, as shown in the left panel of Fig.~\ref{fig_rspee_fqh}.  We decompose the polynomial part of $\Psi_\mathrm{Laughlin}^{1/m}(z_1,...,z_N)$ as a linear combination of Slater determinants \cite{D1993, Sch1993, DiF1994} using an efficient recursive algorithm \cite{BR2009}. The Schmidt decomposition of the state is performed numerically, in analogy with the method for a single Slater determinant discussed around Eq.~(\ref{Eq:LLLd}). We obtain the entanglement entropy for $\Psi_\mathrm{Laughlin}^{1/3}(z_1,...,z_N)$, shown in the right panel of Fig.~\ref{fig_rspee_fqh} as a function of fermion number $N$. The partial sum of the charge cumulant series (as well as $C_2$) shows a clear discrepancy with the entropy. We find similar results if instead of the real space cut we use an orbital cut.

In \ref{Ap:FQHE} we prove that the entanglement entropy can still be obtained from charge cumulants in the simple case of a 2 fermion Laughlin state. We also provide an additional 3 particle counterexample to exemplify how the identity of Eq.~(\ref{Eq:Sse}) fails in general.

\section{One-dimensional effective models of quantum Hall edges}
\label{Sec:1DModels}
Having studied bulk quantum Hall states in the previous section, we turn our attention to the edge degrees of freedom in quantum Hall systems. We first consider a subsystem of length $\ell$ of a one-dimensional edge, where the entanglement spectrum follows solely from bipartite charge fluctuations, thus recovering in the free fermion limit results previously derived in conformal field theory. Next, we evaluate charge fluctuations induced in one half of a quantum Hall sample which has been pinched off by a gating voltage so as to allow weak tunneling between its two halves, which is a problem studied early on by Kane and Fisher \cite{KF1992} and also by Giamarchi and Schulz in a related situation \cite{GiamarchiSchulz}. The case of perfect transmission has been also analyzed by Hsu, Grosfeld and Fradkin in the context of bipartite entanglement \cite{HGF2009}. We calculate charge fluctuations as a function of observation time and recover the entanglement entropies to lowest order in the weak tunneling, as well as some properties of the ES.

\subsection{Critical theories of one-dimensional fermions}
\label{Sec:Gaussian}
We begin by obtaining properties of the entanglement spectrum of critical one-dimensional fermion systems from fluctuations of charge in a contiguous subinterval. We consider the edge subsystem of a quantum Hall sample, following Wen \cite{Wen1992}. We model the edge as a Luttinger liquid, with the Hamiltonian \cite{HaldaneL,G2003}
\begin{equation}
\label{Eq:hll}
H = \frac{u}{2\pi} \int_0^L dx \left[ K (\nabla \theta)^2 + \frac{1}{K} (\nabla \phi)^2  \right].
\end{equation}
Here $u$ is the velocity of excitations, $L$ is the system size, while the Luttinger liquid parameter $K$ accounts for the interaction: $K=1$ corresponds to the noninteracting case, while $K<1$ to repulsion between fermions. In general, $K=\nu$ for the fractional quantum Hall state at filling $\nu$ \cite{Wen1992}. The canonically conjugate fields obey the standard commutation relation $[\phi(x),\nabla \theta(y)] = i \pi \delta(x-y)$. The density of particles can be expressed as $\rho(x) = \rho_0 - \nabla \phi(x)/\pi$.

The charge (i.e., the particle number) enclosed in the subsystem of the length $\ell$ is
\begin{equation}\label{Eq:C2}
N(\ell)=\int_0^\ell dx \rho(x) = \rho_0\ell-\frac{1}{\pi} \left[ \phi(\ell) - \phi(0) \right].
\end{equation}
The last expression enables us to find the second cumulant, which at zero temperature takes the form \cite{SRL2010}
\begin{equation}
C_2=\langle N(\ell)^2\rangle- \langle N(\ell)\rangle^2  = \frac{K}{\pi^2} \log\left(\frac{L}{\pi a}\sin\left(\frac{\pi\ell}{L}\right)\right).
\end{equation}
Here $a$ is a small distance cutoff, while we assume $\ell \gg a$. Since the theory in Eq.~(\ref{Eq:hll}) is gaussian, density fluctuations are determined by $C_2$ and all higher cumulants vanish.

In the noninteracting case, knowledge of $C_2$ enables us to find the R\'enyi entropies, which follow from Eqs.~(\ref{Eq:Sse}) and~(\ref{Eq:Snse}).  
Using $K=1$ for free fermions, we obtain
\begin{equation}
\label{Eq:Snell}
S_n(\ell)=\frac{1}{6}\left(1+\frac{1}{n}\right) \log\left(\frac{L}{\pi a}\sin\left(\frac{\pi\ell}{L}\right)\right),
\end{equation}
while the entanglement entropy is simply obtained as $\lim_{n\to 1}S_n( \ell )$. In the limit $L\to\infty$, this result agrees with the general result for the entanglement entropy in a 1+1-dimensional conformal field theory, $\frac{c}{6} \left( 1 + \frac{1}{n} \right) \log\frac{\ell}{a}$. The central charge, $c$, is equal to unity for noninteracting fermions \cite{CC2009}.

One can now calculate the spectral function $P(\lambda)$ from the sequence $S_n$. As shown in Ref. \cite{CL2008}, substituting Eqs.~(\ref{Eq:Snell}) and~(\ref{Eq:Rn}) in Eq.~(\ref{Eq:fz}), and after evaluating the discontinuity on the real axis, the spectral function takes the form
\begin{equation}
\label{Eq:1dspectrum}
P(\lambda) = \delta(\lambda_{\mathrm{max}} - \lambda) + \frac{b \Theta(\lambda_{\mathrm{max}} - \lambda)}{ \lambda \sqrt{b \log (\lambda_{\mathrm{max}} / \lambda)}} \cdot I_1 \left( 2 \sqrt{b \log(\lambda_{\mathrm{max}}/\lambda)}\right).
\end{equation}
Here and in the following, the functions $I_j$ are the modified Bessel functions of the first kind \cite{GR}, and $\Theta$ is the Heaviside step function. The entanglement spectrum has a non-degenerate largest eigenvalue $\lambda_{\mathrm{max}} = \exp( - b)$, with $b = \lim_{n \rightarrow \infty} S_n(\ell)$. Then the counting function introduced in Eq.~(\ref{Eq:nlambda}) follows easily from the above, $n(\lambda) \equiv \int_\lambda^{\lambda_{\mathrm{max}}} d\Lambda P(\Lambda) =  I_0 ( 2 \sqrt{b \log \lambda_\mathrm{max}/\lambda})$. This analytical form agrees well with numerical results for interacting models at and near criticality \cite{PM2009}.

\subsection{Bipartite fluctuations in time at a weak link}  
\label{Sec:WLt}
In this subsection we consider bipartite charge fluctuations in a Hall sample to which a strong gating voltage $V_g$ is applied as shown in Fig.~\ref{fig_edge_wl}A. The gating voltage allows tunneling between the edge degrees of freedom of the two resulting subsystems. Tunneling between the bulk degrees of freedom is suppressed. We will refer to this regime as a ``weak link''. We use the current noise at the weak link \cite{KF1992,CFW1995,LS1999,BB1999} to obtain the charge fluctuations and from them the entanglement and R\'enyi entropies to lowest order in perturbation theory in the coupling. We note that the problem of quantum entanglement of Luttinger liquids coupled by a quantum impurity has attracted interest recently \cite{Vasseur, Bayat}. 

Let us assume that the weak link is positioned at the middle of the sample, such that $\ell_\mathrm{wl} = L / 2$ [see Fig.~\ref{fig_edge_wl}]. We denote by $J$ the kinetic energy scale in the sample. The weak link term is proportional to $f J$, where $0 < f \ll 1$. The particles tunneling at the weak link are electrons, even in the presence of weak interactions. The corresponding continuum Hamiltonian corresponds of two open boundary Luttinger liquids, denoted by $\alpha=1,2$, coupled by a tunneling term:
\begin{eqnarray}
\label{Eq:H0Vt}
H_0 + V(t) &=& \frac{u}{2\pi} \sum_{\alpha=1}^2  \int_0^\frac{L}{2} dx \left[ K (\nabla \theta_\alpha)^2 + \frac{1}{K} (\nabla \phi_\alpha)^2  \right] \nonumber \\
&&\;\;\;- \frac{f J}{\pi} \cos [ \theta_1(0) -\theta_2(0) + a(t) ]. 
\end{eqnarray}
The term $H_0$ in the first row of Eq.~(\ref{Eq:H0Vt}) describes spinless fermions with a generic repulsive interaction ($K<1$). We are neglecting sine-Gordon terms in $H_0$.  The second row of Eq.~(\ref{Eq:H0Vt}) contains a boundary sine-Gordon term $V(t)$. The time-dependent gauge field $a(t)$ produces the bias voltage $V_b = \partial_t a(t)$. We will consider here the limit $L \to \infty$, when the two subsystems are semi-infinite.

We evaluate cumulants of the charge operator $Q_1(t)$ in subsystem 1 to lowest order in $f$ from the current noise at the weak link $\langle I (t_1) I(t_2) \rangle_\textit{conn.}$ \cite{CFW1995}. We summarize the calculation here for completeness. The Heisenberg picture current operator is  $I(t) = \left(f J / \pi \right) \sin \left[ \theta_1(t) - \theta_2(t) + a(t) \right]$. Note that $I(t) \equiv 2 \partial_t  Q_1(t)$  follows from conservation of charge. We perform standard perturbation theory in powers of $V(t)$ \cite{RS} starting from $\langle I(t) \rangle =\langle T_{c} e^{- i \int_c dt' V_0(t') } I_0(t) \rangle_0 /  \langle T_{c} e^{- i \int_c dt' V_0(t')  } \rangle_0.$ The subscript 0 for an operator denotes the interaction picture with respect to $H_0$, and $\langle O \rangle_0$ denotes the expectation value of an operator $O$ with respect to $H_0$. $T_c$ is the operator that orders on the contour $c$, which passes from $t_0$ to $t$ to $t_0$ to $t_0-i\beta$. $\beta$ is the inverse temperature.

The lowest order contribution to the current at the weak link is quadratic in $f$ 
\begin{eqnarray}
\langle I (V) \rangle \propto \tau_c f^2 J^2  (\tau_c V)^{\frac{2}{K}-1}, 
\end{eqnarray}
which is in agreement with the similar calculation of Kane and Fisher \cite{KF1992}. We introduced the short time cutoff, $\tau_c \equiv a / u$, which regularizes the correlation function at short separations [see Eq.~(\ref{Eq:ExpCo}) below]. We remark that in the noninteracting case the current voltage characteristic is linear, whereas in the presence of repulsive interactions $K < 1$ low bias voltage transport is suppressed. The second cumulant is, to quadratic order, 
\begin{equation}
\langle I(t_1) I(t_2) \rangle = \frac{f^2 J^2}{2\pi^2} \cos \left( V( t_1 - t_2 ) \right) \left\langle  e^{i \sqrt{2} \left[\theta_-(t_1) - \theta_-(t_2) \right] }  \right\rangle_0,
\end{equation}
where $\theta_-(x,t) \equiv \frac{1}{\sqrt{2}} \left[\theta_1(x,t) - \theta_2(x,t)\right]$. The correlation function at the open boundary $x=0$ is
\begin{equation}
\label{Eq:ExpCo}
\left\langle  e^{i \sqrt{2} \left[\theta_-(t_1) - \theta_-(t_2) \right] }  \right\rangle_0 = \left[ \frac{\tau_c/\beta}{ \sinh( \frac{t_1-t_2}{\beta} )} \right]^{\frac{2}{K}}.
\end{equation}
At zero temperature and zero bias, we express the equilibrium charge fluctuations in subsystem 1 as the double time integral
\begin{equation}
\label{Eq:C2t}
C_2( t ) = \langle Q_1^2(t) \rangle = \frac{f^2 J^2}{4 \pi^2} \int_0^t \int_0^t dt_1 dt_2 \left[\frac{\tau_c}{t_1 - t_2 + i \tau_c} \right]^\frac{2}{K} + O( f^4 ).
\end{equation}
The small time cutoff $\tau_c$ in the denominator regulates the integral at vanishing time separations. 

We focus on the free fermion case $K=1$, where charge fluctuations grow logarithmically with time
\begin{equation}
\label{Eq:C2tFree}
C_2( t ) = 
\frac{ \mathcal{ T } }{ 2 \pi^2 } \log \frac{t}{\tau_c}.
\end{equation}
The dimensionless $\mathcal{T} \equiv f^2 | J \tau_c|^2$ is the transmission coefficient through the impurity at the origin. $C_2$ is of order $f^2$ is dominant compared to all higher cumulants $C_{2n} \propto f^{2n}$. In the weak $f$ limit, we therefore truncate the series of Eqs. (\ref{Eq:Sse}) and (\ref{Eq:Snse}) after $C_2$. Then, at the lowest order in perturbation theory we find
\begin{eqnarray}
\label{Eq:St}
S(t) =  \frac{ 1 }{ 3 } \frac{ \mathcal{ T } }{ 2 } \log\frac{ t }{ \tau_c }, \;\; S_n(t) = \frac{ 1 }{ 6 } \left( 1 + \frac{ 1 }{ n } \right) \frac{ \mathcal{ T } }{ 2 } \log\frac{ t }{ \tau_c }.
\end{eqnarray}
The logarithmic dependence is consistent with previous results for free fermions at unit transmission \cite{CC2009,Setal,KL2009,HGF2009}.  The entropy $S(t)$ is also consistent with the results found in Ref. \cite{SetalR}. Moreover, $S_n$ for a subsystem bounded by the weak link is expected to vary as $S_n = \kappa_n \log \ell_\mathrm{wl}$ with a nonuniversal coefficient $\kappa_n \propto f^2$ for small $f \ll 1$ \cite{PE2012}.

The largest eigenvalue in the entanglement spectrum is $\lambda_{\mathrm{max}} = \exp( - b)$, where $b = \lim_{n \to \infty} S_n ( t )$. It has a power law decay in time $\lambda_\mathrm{max} = \left(  t / \tau_c  \right)^{ -  \pi^2 \mathcal{ T } / 12  }$. We note that $\lambda_\mathrm{max} = 1$ at small times $ t \to \tau_c $, and $\lambda_\mathrm{max}$ vanishes at large observation time $t \to \infty$.  $\lambda_\mathrm{max}$ parametrizes the spectral function $P(\lambda)$ derived from Eq.~(\ref{Eq:fz}). As in the case of Eq.~(\ref{Eq:1dspectrum}) in the previous subsection, the expressions of Eq.~(\ref{Eq:St}) allow one to calculate the spectral function $P(\lambda)$ in closed form \cite{CL2008}. The counting function is time dependent $n(\lambda; t) = I_0 \left( 2 \sqrt{b(t) \log \lambda_\mathrm{max}(t)/\lambda} \right)$.

We remark that a power law for the transmission coefficient $\mathcal{T}$ is consistent with an intuitive argument \cite{FisherGlazman} based on Fermi's Golden Rule. In the presence of interactions, effectively free fermions with an interaction-renormalized density of states tunnel at the weak link. On energy scales of the order $\hbar/t$, $\mathcal{T} \propto | \rho( E \sim \hbar / t )  |^2 \propto \left(  t / \tau_c \right)^{2\left(-1/K + 1 \right)}$. This is the same as one obtains from the double integration in  Eq.~(\ref{Eq:C2t}).  In principle, the current fluctuations can be probed with present technology \cite{Glattli,Picciotto}.

\subsection{Numerical tests using quantum spin chains}
In this subsection, we study microscopic models which map in the continuum limit to the Luttinger liquid theory presented in the Subsec.~\ref{Sec:WLt}. We numerically probe the spatial dependence of $S$ and $C_2$ to check the analytical result of Eq.~(\ref{Eq:C2t}). We are considering two microscopic models: (i) The XX quantum spin chain, which maps to free fermions. We are using the density matrix renormalization group method (DMRG) \cite{White} for convenience, where the determination of cumulants and entropies is routine. (ii) The Heisenberg limit of a quantum spin chain, which maps to interacting fermions with nearest neighbor repulsion. We obtain $C_2$, and its dependence on the subsystem length, using quantum Monte Carlo at $T=0$ in the SSE framework \cite{Sandvik02}. 

\begin{figure}[t!]
\includegraphics[width=\linewidth]{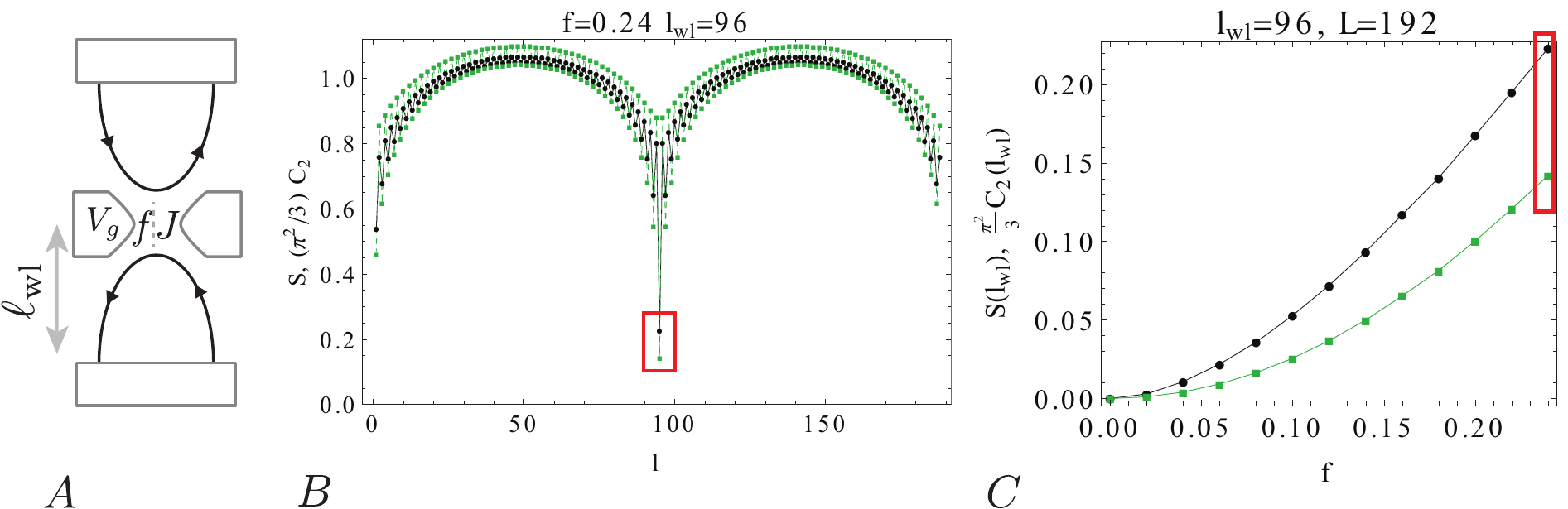}
\caption{\label{fig_edge_wl} \textbf{A:} Sketch of quantum Hall bar with a quantum point contact, modeled as a weak link. The strong gate voltage $V_g$ allows only weak tunneling of electrons between subsystems. \textbf{B:} DMRG results for the entanglement entropy $S$ (black circles) and $\left( \pi^2 / 3 \right) C_2$ (green squares) versus position $\ell$ on the chain if the weak link is in the middle $\ell_\mathrm{wl} = L / 2$. An XX chain of $L=192$ sites was considered (Ising term $J_z = 0$). \textbf{C:} For the same system, entropy $S(\ell_\mathrm{wl})$ (black circles) and $\left( \pi^2 / 3 \right) C_2(\ell_\mathrm{wl})$ (green squares) at the position of the weak link versus weak link strength $f$. The rectangle marks the corresponding values for $f=0.24$ plotted in panel \textbf{B}.}
\end{figure} 

The quantum spin chain with a weak link residing on the $\ell_\mathrm{wl}^\textit{th}$ bond with open boundary conditions is represented by the Hamiltonian:
\begin{eqnarray}
\label{Eq:Hspinwl}
H = &&\sum_{i=1}^{\ell_{\mathrm{wl}} - 1} \left[ J ( S_i^x S_{i+1}^x + S_i^y S_{i+1}^y ) + J_z S_i^z S_{i+1}^z \right] \nonumber \\ && + f \left[ J ( S_{\ell_{\mathrm{wl}}}^x S_{\ell_{\mathrm{wl}}+1}^x + S_{\ell_\mathrm{wl}}^y S_{\ell_\mathrm{wl}+1}^y ) + J_z S_{\ell_\mathrm{wl}}^z S_{\ell_\mathrm{wl}+1}^z \right] \nonumber \\ && + \sum_{i=\ell_\mathrm{wl}+1}^{L-1} \left[ J ( S_i^x S_{i+1}^x + S_i^y S_{i+1}^y ) + J_z S_i^z S_{i+1}^z \right].
\end{eqnarray}
The Hamiltonian of Eq.~(\ref{Eq:Hspinwl}) maps to an interacting fermion Hamiltonian via the Jordan-Wigner transformation. The standard treatment is detailed in \ref{Ap:XY}. If $J_z = 0$, this reduces to a problem of free fermions. We are interested in the $S_z = 0$ sector of the many body Hilbert space, which corresponds to an average density of one spinless fermion for every two sites, i.e. half filling.

Consider first noninteracting fermions. In Subsec.~\ref{Sec:WLt} we derived expressions for $C_2(t)$ and $S(t)$ in a subsystem separated from the environment by a weak link of strength $f$, Eqs. (\ref{Eq:C2tFree}) and (\ref{Eq:St}). As functions of the observation time $t$, we found logarithmic dependences for free fermions $\propto f^2 J^2 \log( t / \tau_c )$. We show now that with a weak link placed at $\ell_{\mathrm{wl}}$ in an XX chain, $S(\ell_{\mathrm{wl}})$ and $C_2(\ell_{\mathrm{wl}})$ are $\propto f^2 J^2 \log(\ell_{\mathrm{wl}}/a)$, with $f \in [0,1]$ being the strength of the link.  We compute entropies and cumulants with DMRG \cite{White}. In DMRG it is possible to obtain a truncation of the reduced density matrix $\rho_A$ for the each subsystem entailing sites $\{0,...,\ell\}$, from which entropies $S_n(\ell)$ and charge cumulants $C_n(\ell)$ are obtained without additional computational cost. In the DMRG routine, we maintained $D = 400$ Schmidt states for the representation of the reduced density matrix, and performed $5$ DMRG sweeps.  

Introducing the compactified length $\ell^{c}_\mathrm{wl} = \frac{L}{\pi} \sin \frac{\ell_\mathrm{wl} \pi}{L}$, we have $S(\ell_\mathrm{wl}), C_2(\ell_\mathrm{wl}) \approx  (b + c \log \ell^c_\mathrm{wl}) f^2$ for free fermions. In Fig.~\ref{fig_edge_wl}, fit results are $b = 1.07 \pm 0.03$ and $c = 2.37 \pm 0.1$ for $S$ and $b = -0.01 \pm 0.07$ and $c = 2.30 \pm 0.02$ for $C_2$. Consistency with the analytical results in the time domain of the previous subsection, Eqs. (\ref{Eq:St}) and (\ref{Eq:C2tFree}), follows from arguments in Ref. \cite{KL2009_2}; the entanglement within a time interval $t$ can be interpreted as the entanglement of a subregion of length $\ell = u t$ with the rest in a stationary free fermion system, where $u$ is the Fermi velocity of the fermions at half filling.

\begin{figure}[t!]
\includegraphics[width=\linewidth]{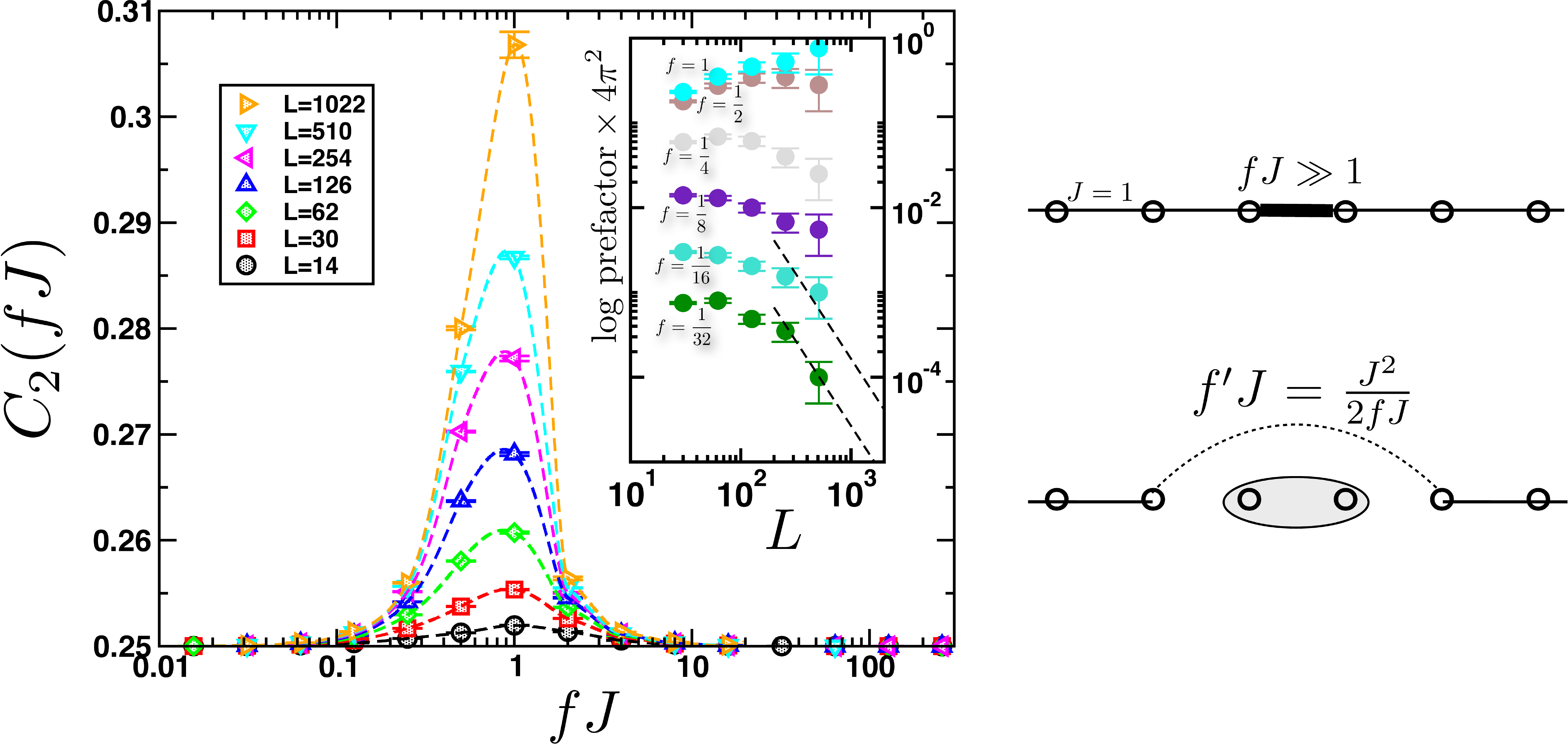}
\caption{\label{fig_wl_su2} $C_2$ as a function of $f$, with $C_2 = \frac{1}{4}$ in both limits $f \ll 1$ and $f \gg 1$. The inset shows the prefactor of the logarithm $\log(L)$ in $C_2$ as a function of $L$ and a few values of $f \ll 1$. The resulting fits are explained in the text.}
\end{figure}

Secondly, we focus on the Heisenberg chain, where $J_z = J$. This corresponds to fermions with nearest neighbor repulsion. Numerical results from quantum Monte Carlo are presented in Fig.~\ref{fig_wl_su2}. The weak link is at the middle of the chain, such that $\ell_\mathrm{wl} = L/2$, and $L/2$ is an odd integer. $C_2$ approaches the value $\frac{1}{4}$ in both limits $f \ll 1$ and $f \gg 1$ \cite{Footnote1}. The latter case maps to the former: a singlet forms on the $\ell_\mathrm{wl}^\textit{th}$ bond, and a new tunneling term is obtained from second order perturbation theory in $J \ll fJ$, yielding $f' J = \frac{J^2}{2 f J}$ (see Fig.~\ref{fig_wl_su2}). Fig.~\ref{fig_wl_su2} shows $C_2$ as a function of $f$. We note that $C_2 = \frac{1}{4}$ in both limits $f \ll 1$ and $f \gg 1$ in agreement with renormalization group arguments. Agreement with the Kane Fisher result \cite{FisherGlazman} can be verified by fitting the prefactor of the logarithm in $C_2 ( L / 2 ) \propto L^{2(1-1/K)} \log( L )$ at large $L$ and for $f \ll 1$. The fit is consistent with $K = 1/2$ at the Heisenberg point. We recall that logarithmic corrections occur in certain quantities, such as correlation functions, for the point $K=1/2$, since the umklapp term becomes important.

\section{Conclusions}
To summarize, based on previous work \cite{Setal, SetalR}, we have revealed a correspondence between the bipartite charge fluctuations and the real space partition entanglement spectrum for the integer quantum Hall state. This correspondence is valid in the limit of large fermion density, but good agreement is found for particle numbers accessible to numerical evaluations of the RSP ES. We have argued that the main properties of the entanglement spectrum can be accessed from a measurement of the second charge cumulant. Secondly, we have discussed the relation between charge fluctuations and R\' enyi entropies in the case of a quantum point contact, which as such models the edge subsystem of a quantum Hall sample strongly gated so as to allow weak tunneling between two subsystems. Recently, the current noise of excitations generated by voltage pulses has been measured at a quantum point contact \cite{Dubois}. We have also addressed the issue of extending the free fermion results to fractional quantum Hall trial states. Looking at the Laughlin state at filling $\nu=\frac{1}{3}$, we have shown that this is not generally possible, except for the two fermion case. This analysis could be extended to other systems, for example to fermions with a BCS interaction \cite{KlichI}.

\section*{Acknowledgements} 
It is a pleasure to thank I. Affleck, B. A. Bernevig, S. Bose, J. Dubail, E. Fradkin, J. Gabelli, C. Glattli, F. D. M. Haldane, A. L\"auchli, L. Levitov, T. Liu, M. Pepper, B. Reulet, A. Sterdyniak and A. Turner, for interesting discussions related to this work. This work has also benefitted from discussions at the CIFAR meetings in Canada. This project is supported by the LABEX PALM at Paris-Saclay, project Quantum-Dyna. SR is supported by the DFG through FOR 960 and by the Helmholtz association through VI-521. CF is supported by QSIT and the Swiss NSF. IK is supported by the NSF CAREER grant DMR-0956053. NL is supported by the French ANR program ANR-11-IS04-005-01. NR was supported by the Princeton Global Scholarship.

\begin{appendix}
\section{Numerical study of entanglement measures in XXZ spin chains}
\label{Ap:XY}
In this appendix we study the breakdown of the free fermion result Eq.~(\ref{Eq:Sse}) with interactions. For this purpose, we compute cumulants, entropies, and the entanglement spectrum in the quantum XXZ chain 
\begin{equation}
\label{Eq:Hspin}
H = \sum_i \left[ J ( S_i^x S_{i+1}^x + S_i^y S_{i+1}^y ) + J_z S_i^z S_{i+1}^z \right].
\end{equation}
The index $i$ runs between $1$ and $L-1$ for open boundary conditions (OBC) and between $1$ and $L$, with $L + 1 \equiv 1$, for periodic boundary conditions (PBC). The standard Jordan-Wigner mapping of spin $1/2$ operators to fermionic operators \cite{G2003} turns Eq.~(\ref{Eq:Hspin}) into the Hamiltonian of lattice fermions with next-neighbor interaction, 
\begin{equation}
\label{Eq:Hfermi}
H = \sum_i \left[ \frac{J}{2} \left( c_i^\dagger c_{i+1} + \mathrm{H.c.} \right) +  J_z \left( c_i^\dagger c_i - \frac{1}{2} \right) \left( c_{i+1}^\dagger c_{i+1} - \frac{1}{2} \right) \right].
\end{equation}
Total spin $S_z = 0$ corresponds to the half-filled fermion system: total fermion number $\frac{L}{2}$ on $L$ sites when $L$ is even. The Hamiltonian in Eq.~(\ref{Eq:Hspin}) can be routinely studied using the density matrix renormalization group method (DMRG) \cite{White}. 

The sequence of charge cumulants $C_n(\ell)$ is determined in DMRG as follows (cf. Appendix of \cite{Setal}): the reduced density matrix is stored in block-diagonal form with respect to eigenvalues of the total spin operator $S^z (\ell) = \sum_{i=1}^\ell S_i^z$, equivalently total fermion number $N(\ell) = S^z(\ell) + \frac{\ell}{2}$. Denote a given block of the reduced density matrix by $\alpha$, the corresponding eigenvalue of $S^z(\ell)$ by $m^\alpha(\ell)$, and the eigenvalues of the reduced density matrix by $w_j^\alpha$, with $j$ ranging from $1$ to the subspace dimension. The $p^\textit{th}$ moment of fermion density is
\begin{equation}
\langle N^p(\ell) \rangle = \sum_\alpha \left( \sum_j w_j^\alpha \right)  \left( m^\alpha(\ell) + \frac{\ell}{2} \right)^p.
\end{equation}
Cumulants $C_n(\ell)$ are efficiently evaluated from moments using recursion relations \cite{Smith}.

\begin{figure}[t!]
\includegraphics[width=\linewidth]{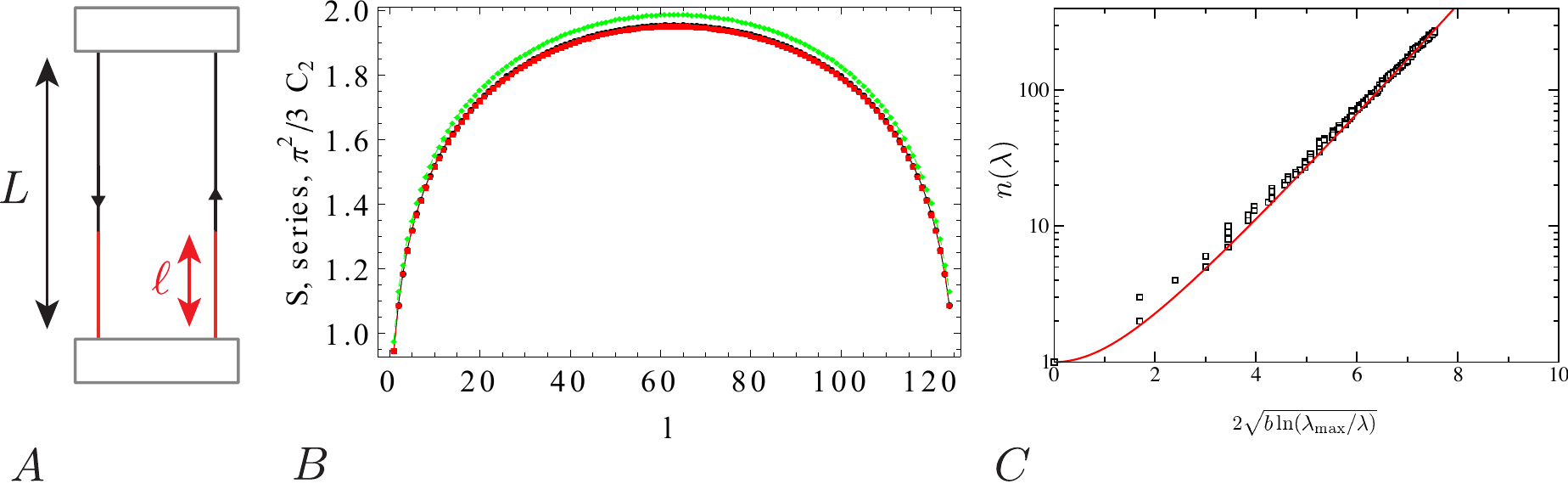}
\caption{\label{supfig_edge_l} The one-dimensional edge of a quantum Hall sample is separated into subsystems $(0,\ell)$ and $(\ell,L)$. \textbf{A:} Rectangular geometry, with a chiral mode on each edge. \textbf{B:}  DMRG results for a periodic 128 site XX chain at total spin $S_z = 0$; the entanglement entropy (black circles), series results (red squares) and $\left( \pi^2 / 3 \right) C_2$ (green rhombi);  \textbf{C:} ES for a real space cut at $\ell = 1000$ in a 2000-site XX chain with open boundary conditions. The solid \textit{red line} shows the counting function $n(\lambda)$ of Eq.~(\ref{Eq:1dspectrum}) and the \textit{black squares} show the ES from DMRG.}
\end{figure}

For free fermions $J_z = 0$, Eq.~(\ref{Eq:Sse}) containing the series result for $S(\ell)$ is easily confirmed numerically as shown in the main plot of Fig. \ref{supfig_edge_l}B. Note that exact agreement between $S$ and $(\pi^2 / 3) C_2$ is expected in the infinite chain limit $L \to \infty$. The minor deviation of $(\pi^2 / 3) C_2$ from $S$ in Fig.~\ref{supfig_edge_l}B is due to finite size effects. Nevertheless, we sum the first $32$ terms of the series Eq.~(\ref{Eq:Sse}) containing higher order cumulants $C_4$, $C_6$ etc. and find very good agreement between the partial sum of the series and the entropy.

We remove finite size effects by appropriate fitting of $S$ and $C_2$ as functions of subsystem length $\ell$. We exemplify this in the XXZ regime $J > J_z > 0$. While Eq.~(\ref{Eq:Sse}) should not hold, there is an exact relation between $S$ and $C_2$ as $L \to \infty$.  
We fit the leading order logarithmic contribution to $S$ and $C_2$ for $0 \ll \ell < L$. Upon taking the ratio of the coefficients of the logarithm, we find, as expected from analytical results \cite{SRL2010, Setal}
\begin{equation}
\label{Eq:SC2}
\frac{S(\ell)}{C_2(\ell)} \sim \frac{\pi^2}{3} \frac{1}{K},\;\mathrm{with}\; \frac{J_z}{J} = - \cos\left( \frac{\pi}{2K} \right)
\end{equation}
for large $\ell$. Here, $K$ denotes the Luttinger parameter in the bosonized theory for the XXZ spin chain \cite{G2003}. $K = 1$ for free fermions and $\frac{1}{2}$ at the Heisenberg point $J_z = J$. This is consistent with $S/C_2 \sim 2 \pi / 3 \arccos\left( - J_z / J \right)$ at large system length $L \to \infty$. Relation Eq.~(\ref{Eq:SC2}) parametrizes the breakdown of the free fermion result as interactions increase.

We finally obtain the ES counting function $n(\lambda)$ introduced in Eq.~(\ref{Eq:nlambda}) of the main text. The single parameter determining $n(\lambda)$ is $b = - \log \lambda_\mathrm{max}$. We consider a 2000-site open boundary XX chain and a real space cut at $\ell = 1000$. We find good agreement between $n(\lambda) = I_0 \left( 2 \sqrt{b \log \lambda_\mathrm{max}/\lambda} \right)$ \cite{CL2008} and the numerically computed counting function in Fig.~\ref{supfig_edge_l}C. Note that this agreement is expected to persist in the entire critical region $1 \geq J_z  > -1$. Highly accurate results \cite{PM2009} have confirmed the form Eq.~(\ref{Eq:1dspectrum}) in quantum lattice models at or near a critical point.


\section{RSP ES for bulk IQHE}
\label{Ap:esiqhe}
In this appendix we discuss the regime of validity of the approximation involved in obtaining $n(\lambda)$ for the RSP ES of the fully filled LLL discussed in Sec.~\ref{Sec:bulkIQHE}. Empirically, we find that the approximation is justified if the following condition on the geometry is respected: 

\textit{The surface of the subsystem $A$ and the surface of its complement $B$ with respect to the area of the quantum Hall disk have to accommodate a large number of orbitals $\phi_m(z)$}. 

Note that the subsystem $A$ was defined as an annular subdomain of the disk, $A = \{\; z\;  |\;\; |z| \in \left[ r_1, r_2 \right]\;\}$. The condition that many orbitals are supported in $A$ is equivalent to the condition that $m_1 \ll m_2$, where $m_1 = \frac{r_1^2}{2 l^2}$ and $m_2 = \frac{ r_2^2 }{ 2 l^2 }$ are the angular momenta of LLL wavefunctions which are strongly localized around $r_1$ and $r_2$ respectively. Noting that the surface area of the annular subsystem is in fact $\mathcal{A} = \pi( r_2^2 - r_1^2 ) = 2 (m_2 - m_1) \pi l^2$, this condition is equivalent to requiring that the subsystem surface is large $\mathcal{A} \gg 2 \pi l^2$. Experimentally, this condition can be fulfilled by decreasing the magnetic length. Since the magnetic length in SI units is $l = \sqrt{\frac{\hbar c}{e B}}$, increasing the magnetic field accommodates more orbitals in the subsystem $A$, and the number of orbitals enclosed in the subsystem is approximately $\mathcal{A}/(2 \pi l^2)$. In addition, particle-hole symmetry requires that the area $\mathcal{B}$ of the complementary subsystem $B$ obey a similar condition $\mathcal{B} \gg 2 \pi l^2$. 

\begin{figure}[t!]
\includegraphics[width=\linewidth]{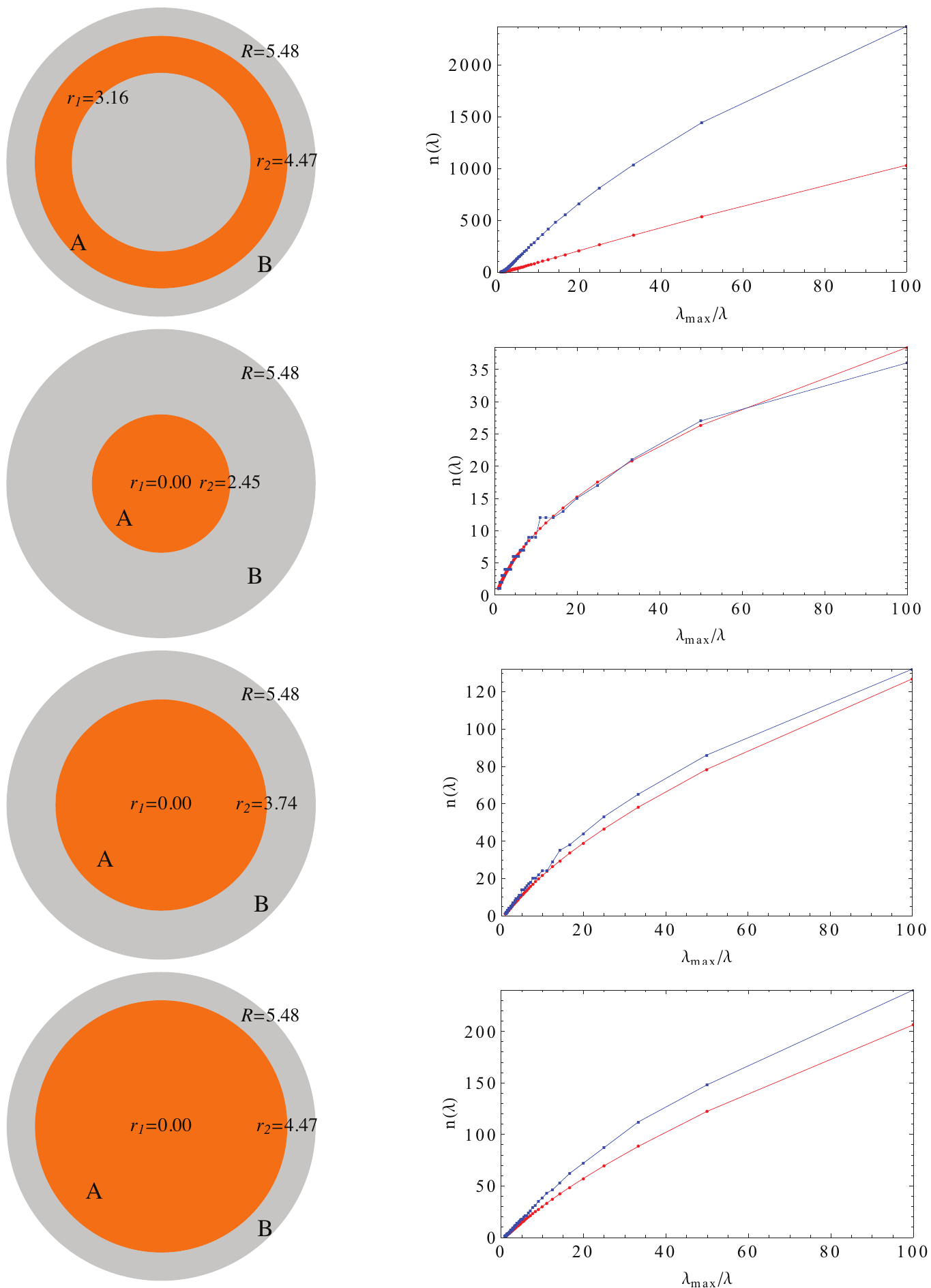}
\caption{\label{supfig_rspes_iqh}Entanglement spectrum counting function obtained from $C_2$ (red circles) and the exact counting function (blue squares). Radii in the left panels are in units of the magnetic length $l$. The disk radius is $R = \sqrt{30} l$. \textbf{$1^\mathrm{st}$ row:} $r_1 = \sqrt{10} l$, $r_2 = \sqrt{20} l$, about 10 orbitals in subsystem $B$ and 5 in $A$. \textbf{$2^\mathrm{nd}$ row:} $r_1 = 0$, $r_2 = \sqrt{6} l$, 12 orbitals in $B$ and 3 in $A$. \textbf{$3^\mathrm{rd}$ row:} $r_1 = 0$, $r_2 = \sqrt{14} l$, 8 orbitals in $B$ and 7 in $A$. \textbf{$4^\mathrm{th}$ row:} $r_1 = 0$, $r_2 = \sqrt{20}l$, about 10 orbitals in $A$ and 5 in $B$. We find best agreement when the areas $\mathcal{A}$ and $\mathcal{B}$ of $A$ and $B$ are comparable, and when $A$ and $B$ are connected subdomains of the disk.}
\end{figure}

The condition above leads to a simpler condition on the radii: $R \gg r_2 \gg r_1 \gg l$. These inequalities are also sufficient for the entropies and cumulants to satisfy boundary laws. Additionally, we found that our approximation of the RSP ES spectral function from charge fluctuations works better if we require that the entanglement surface is a single circle: $R = r_2 > r_1 \gg l$ or $R \gg r_2 > r_1 = 0$. This is equivalent to requiring that both $A$ and $B$ are connected. We show numerical checks of these qualitative observations in Fig.~\ref{supfig_rspes_iqh}.

\section{Cumulant series for few particle Laughlin states}
\label{Ap:FQHE}
In this appendix we prove that the entanglement entropy arises from cumulants via Eq.~(\ref{Eq:Sse}) in the simple case of a 2 fermion Laughlin state (this approach is not possible for more than $2$ particles). For simplicity, we will take here the magnetic length $l$ equal to unity.

\subsection{2-fermion Laughlin state}
\label{Subsec:2part}
For the two fermion state $\Psi^{1/3}_\mathrm{Laughlin}(z_1,z_2)$ we can build a correspondence between entropies and charge cumulants. In this case, $\rho_A$ is sum of contributions from each individual Slater determinant in the expansion of $\Psi_\mathrm{Laughlin}$. The series of cumulants are then applicable for each Slater determinant. 

The 2-fermion Laughlin state at $\nu = 1/3$ has the following decomposition in terms of Slater determinants
\begin{equation}
\label{Eq:L3N2}
\Psi^{1/3}_\mathrm{Laughlin}(z_1, z_2) \propto 
\left|\begin{array}{cc}
z_1 & 1 \\
z_2 & 1 \\
\end{array}\right|^3  = 
\left|\begin{array}{cc}
z_1^3 & 1 \\
z_2^3 & 1 \\
\end{array}\right|
- 
3\left|\begin{array}{cc}
z_1^2 & z_1 \\
z_2^2 & z_2 \\
\end{array}\right|.
\end{equation}
For simplicity, assume that $z_1, z_2$ are planar coordinates in a disk sample. The proportionality sign indicates the absence of an overall normalization constant. Let us switch to notations introduced in Sec.~\ref{Sec:bulkIQHE} to perform the Schmidt decomposition. Denoting sets of orbitals $\mathcal{O}_\mathcal{M} = \{0,3\}$ and $\mathcal{O}_\mathcal{N} = \{1,2\}$, Eq.~(\ref{Eq:L3N2}) becomes
\begin{eqnarray}
\label{Eq:phiMN}
| \Psi_\mathrm{Laughlin}^{1/3} \rangle &=&  \mathcal{M}\prod_{m\in \mathcal{O}_\mathcal{M}} (\zeta_m^{1/2} c_{Am}^\dagger + ( 1 - \zeta_m )^{ 1 / 2 } c_{Bm}^\dagger ) | 0 \rangle  \nonumber \\
&& +\mathcal{N} \prod_{n \in \mathcal{O}_\mathcal{N}} (\zeta_n^{1/2} c_{An}^\dagger + ( 1 - \zeta_n )^{ 1 / 2 } c_{Bn}^\dagger ) | 0 \rangle.
\end{eqnarray}
 Up to the overall normalization of $|\Psi_\mathrm{Laughlin}^{1/3} \rangle$, the weight of the first Slater is $\mathcal{M} = 1 \cdot \sqrt{2 \pi 2^3 3!} \cdot \sqrt{2\pi 2^0 0!}$, whereas the weight of the second is $\mathcal{N} = -3 \sqrt{2 \pi 2^2 2!} \sqrt{2 \pi 2^1 1!}$. These follow from the normalization of the single particle orbitals [see Eq.~(\ref{Eq:phim})].

The Schmidt decompositions of the two Slater determinants are performed independently following the recipe of Eq.~(\ref{Eq:LLLd}).
Forming $\rho = | \Psi_\mathrm{Laughlin}^{1/3} \rangle \langle \Psi_\mathrm{Laughlin}^{1/3} |$ and tracing over degrees of freedom in $B$ leads to the reduced density matrix. Since the sets of orbitals $\mathcal{O}_\mathcal{M}$ and $\mathcal{O}_\mathcal{N}$ are disjoint, the reduced density matrix can be written $\rho = \mathcal{M}^2 \rho_\mathcal{M} + \mathcal{N}^2 \rho_\mathcal{N}$. We define $\rho_\mathcal{M} \equiv \sum_i \exp( - \xi_i^\mathcal{M} ) |\psi_{A\mathcal{M}}^i \rangle  \langle \psi_{A\mathcal{M}}^i |$. The indices $i = 1,...,2^2$ denote the possible occupation configurations $\{n_m^i\}$ over the $2$ orbitals in the set $\mathcal{O}_\mathcal{M}$ (analogous definitions hold for $\rho_\mathcal{N}$). Then
\begin{eqnarray}
S &=& - \mathcal{M}^2 \mathrm{Tr} \left[ \rho_\mathcal{M} \log( \rho_\mathcal{M})  \right] - \mathcal{N}^2 \mathrm{Tr} \left[ \rho_\mathcal{N} \log( \rho_\mathcal{N})  \right] \nonumber \\ &&- \mathcal{M}^2 \log \mathcal{M}^2 - \mathcal{N}^2 \log \mathcal{N}^2 .
\end{eqnarray}
Note that each Slater determinant yields a term in the entanglement entropy. By using the cumulant series for each term, we recover the entropy $S$.

\subsection{A 3 particle counterexample}
In general the Slater determinants in the decomposition of $\Psi_\mathrm{Laughlin}^{1/m}$ correspond to non-disjoint orbital sets. Consider however the Laughlin wavefunction of 3 particles, where the decomposition of $\Psi_\mathrm{Laughlin}^{1/3}(z_1, z_2, z_3)$ is \cite{D1993}
\begin{eqnarray}
\left|\begin{array}{ccc}
z_1^2 & z_1 & 1 \\
z_2^2 & z_2 & 1 \\
z_3^2 & z_3 & 1
\end{array}\right|^3  =
\left|\begin{array}{ccc}
z_1^6 & z_1^3 & 1 \\
z_2^6 & z_2^3 & 1 \\
z_3^6 & z_3^3 & 1 
\end{array}\right|
-3 \left|\begin{array}{ccc}
z_1^6 & z_1^2 & z_1 \\
z_2^6 & z_2^2 & z_2 \\
z_3^6 & z_3^2 & z_3 
\end{array}\right| \nonumber \\
- 3\left|\begin{array}{ccc}
z_1^5 & z_1^4 & 1 \\
z_2^5 & z_2^4 & 1 \\
z_3^5 & z_3^4 & 1 
\end{array}\right| 
 + 6\left|\begin{array}{ccc}
z_1^5 & z_1^3 & z_1 \\
z_2^5 & z_2^3 & z_2 \\
z_3^5 & z_3^3 & z_3 
\end{array}\right|
- 15\left|\begin{array}{ccc}
z_1^4 & z_1^3 & z_1^2 \\
z_2^4 & z_2^3 & z_2^2 \\
z_3^4 & z_3^3 & z_3^2 
\end{array}\right|.
\end{eqnarray}
For example, cross terms appear between the first two terms. The orbital sets are $\mathcal{O}_\mathcal{M} = \{ 0, 3, 6 \}$ and $\mathcal{O}_\mathcal{N} = \{ 1, 2, 6 \}$, which have a common orbital $m = 6$. The two Slaters have a nonvanishing overlap on the subregion $A$ coming from the Schmidt basis element corresponding to orbital occupancy $\{n_m\} = [ 0,0,0,0,0,0,1 ]$
\begin{equation}
\left( \prod_{m\in \mathcal{O}_\mathcal{M}} \langle 0 | (c_{Am})^{n_m} \right) \;\left(\prod_{m' \in \mathcal{O}_\mathcal{N}} (c_{A m'}^\dagger)^{n_{m'}} | 0 \rangle \right) = 1. 
\end{equation}
Cross terms of the form $\mathrm{Tr}\left[ \rho_\mathcal{M} \log \rho_\mathcal{N} \right] \neq 0$, appear in the expression of $S$. Therefore the treatment of \ref{Subsec:2part} does not apply in general.

\end{appendix}


\begin{thebibliography}{90}
\bibitem{Footnote1}
In the limit of asymptotically decoupled wires, each of them having an odd number of spins, i.e. a doublet ground state, they eventually couple to form a low energy singlet which contributes 1/4 to C2.
\end{thebibliography}
\end{document}